# α/γ discrimination method for bulky BaF₂ detector used in γ total absorption facility*

ZOU Chong[1,2], ZHANG Qiwei[1], LUAN Guangyuan[1], WU Hongyi[1], LUO Haotian[1], CHEN Xuanbo[1], WANG Xiaoyu[1], HE Guozhu[1], REN Jie[1], HUANG Hanxiong[1], RUAN Xichao[1], BAO Jie[1], ZHU Xinghua[3]

1.Key Laboratory of Nuclear Data, China Institute of Atomic Energy, Beijing 102413, China

2.China Shipbuilding Trading Co., Ltd., Beijing 100044, China

3.Huaneng Shandong Shidao Bay Nuclear Power Co., Ltd., Rongcheng 264312, China

**Abstract**

The gamma-ray total absorption facility (GTAF) composed of 40 BaF₂ detection units is designed to measure the cross section data of neutron radiation capture reaction online, in order to comply with the experimental nuclear data sheet. Since 2019, several daunting experiment results have been analyzed and published, and we have found that one of the most important sources of experimental background is the initial α particles emitted by the BaF₂ crystal, which is the core component of GTAF detection unit.Considering the current industrial manufacturing process capabilities, the impurities of Ra and its compounds cannot be completely removed from the BaF₂. Developing data analysis algorithms to eliminate the influence of alpha particles in experimental data has become a key aspect. In this work, in order to meet the needs of data acquisition, online measurement and analysis of neutron radiation cross section, the GTAF data acquisition system adopts a full waveform acquisition method, which results in a large number of data recorded, transmitted, and stored during experiment, which also affects the uncertainty of the cross-section data. The number of data stored in the online experiment is about 118 MB/s, resulting in a long dead time.Based on the signal waveform characteristics of the BaF₂ detection unit, in order to solve the aforementioned problems, three methods, namely the ratio of fast component to total component, pulse width, and time decay constant, are used to identify and distinguish α particles and γ rays. The quality factor FOM is utilized as an evaluation value and several experiments are conducted using three radioactive sources ($^{22}$Na, $^{137}$C, $^{60}$Co) for verification.Due to the slow components of BaF$_2$ light decay time being about 620 ns, the waveform pulse should essentially return to baseline at approximately 1900 ns to 2000 ns, allowing for the complete waveform of the γ rays signal to be captured at that moment, which may provide the best energy resolution. Therefore, in the online experiment, the integration length for the energy spectrum is chosen to be 2000 ns in this work.The quality factor is 1.19–1.41 from the fast total component ratio (fast component 5 ns, total component 200 ns) method, 0.94–1.04 from the pulse width (10% peak) method, and 0.93–1.07 from the time attenuation constant method. Through the quantitative analysis of quality factor and the comparison of energy spectrum, it is determined that the fast total component ratio method has the best effect and can effectively remove the background of α particles.The next step is to upgrade the online experimental data acquisition system to reduce the quantity of experimental data and the uncertainty of cross section data. The experimental data that need to be recorded should be the crossing threshold time (for the time-of-flight method) and the amplitude integration values of 5 ns after the threshold (for the fast component), 200 ns after the threshold (for the total component), and 2000 ns (for the energy) for each signal waveform, as well as the number of related detection units. The above information should be sufficient to complete online processing of experimental data, including the processing of the α particle background and (n,γ) reaction data. It is estimated that the data acquisition rate of the upgraded system will decrease from 118 MB/s to 24 MB/s, which can significantly reduce the dead time of the data acquisition system, thereby improving the accuracy of cross section data.

**Keywords:** energy resolution, ratio of fast to total component, pulse width, time decay constant

**PACS:**28.20.Np, 07.85.Nc, 01.52.+r

# 1. Introduction

The radiative neutron capture reaction, (n, γ) reaction, is the dominant neutron disappearance reaction from thermal neutron energy region to keV energy region. The cross section data of (n, γ) reaction is widely used in advanced nuclear energy research[1], reactor design[2] and nuclear astrophysics research[3]. The delayed gamma method (activation method) was used to measure the (n, γ) reaction cross section data in the early stage in the world, but it is only suitable for the experimental conditions of monoenergetic neutrons, and it is difficult to distinguish the data of the same product nucleus obtained through different reaction channels. Since the 1960s, the time-of-flight (TOF) prompt gamma method (on-line method) has been used to measure (n, γ) reaction cross section data on large volume liquid scintillation[4], Moxon-Rae[5], NaI (Tl) [6], BGO[7] and other different types of detectors. However, due to the limitations of these detectors, the measured cross section data have large uncertainties. Since the 1990s, several important laboratories in the world, such as ORELA[8] of ORNL in the United States, GELINA[9] of JRC in Europe, n _ TOF[10] of CERN, KURRI-LINAC[11] of Japan and Back-n[12] of CSNS, have successively established $C_6D_6$ detectors to measure the (n, γ) reaction cross section data of stable nuclides. In the 1980s, it was found that $BaF_2$ crystal had the characteristics of high detection efficiency and good time resolution, and began to be made into scintillator detectors for γ-ray measurement. Three sets of γ-total absorption $BaF_2$ detection devices were built in the world, namely, $4\pi BaF_2$ [13] of FZK Karlsruhe in Germany, TAC[14] of CERN in Europe (Switzerland) and DANCE[15,16] of LANL in the United States, which were used to measure the neutron capture cross-sections of fission nuclides, radionuclides, and nuclei that have low sample amounts and small cross-sections.. At present, the measurement of (n, γ) reaction cross section data of fissile nuclides in the range of $A$< 120 and $A$ > 180 has become a hot topic of nuclear data research in the world.

Due to the lack of high-quality detection devices and measurement methods in China, coupled with the barrier protection of key nuclear cross-section data in developed countries, some important fissile nuclides, such as$^{235}$U,$^{239}$Pu,$^{241}$Am, lack of cross-section data or have large differences in data, which can not meet the needs of engineering construction and scientific research. In order to fill this gap, the Key Laboratory of Nuclear Data of China Institute of Atomic Energy (CIAE) has built the first gamma total absorption facility (GTAF) in China. Based on the inverse angle white light neutron source Back-n of China Spallation Neutron Source, some on-line (n, γ) reaction cross section measurements of stable nuclides have been carried out, and good experimental results have been obtained[17–19].

The GTAF consists of 40 $BaF_2$ detection units. By analyzing the data of the on-line experiment, it is found that the α particle contained in the $BaF_2$ detection unit itself is an important source of background, and the key to reduce the uncertainty of the cross section data is to deduct the α particle background to improve the effect background ratio[20–22]. At present, the data acquisition system of GTAF uses the full waveform acquisition mode, and the signal waveforms of all detection units meeting the trigger conditions will be recorded. The advantage of this method is that in the process of off-line data processing, the alpha particle background can be deducted by waveform analysis to obtain the energy information and time information of the signal; The disadvantage is that the amount of data that needs to be collected, transmitted and stored is huge. Under the beam current condition of 140 kW power of China Spallation Neutron Source, the average data volume of GTAF online experiment is 118 MB/s, and the peak data volume can reach 160 MB/s. There are extremely high requirements for the trigger selection and data acquisition of electronics. The dead time of the data acquisition system reaches more than 20%, which directly affects the uncertainty of the cross section data. With the increase of the beam power of the spallation neutron source, the data volume of the online experiment will continue to increase. Therefore, an upgrade of the data acquisition system is planned, aimed at reducing the amount of data the system is required to record.

## 2. GTAF and BaF$_2$ detection unit of γ total absorption facility

BaF$_2$ crystal is an inorganic scintillator[23–26], which has two luminescent components: the fast component has a light decay time of 0.6 ns and a wavelength peak at 225 nm, which determines that the BaF$_2$ crystal has very good time resolution; The optical decay time of the slow component is 620 ns, and the peak wavelength is 325 nm; The neutron sensitivity of BaF$_2$ crystal is low and its refractive index is close to that of glass, which makes it easier for scintillation light to enter the photomultiplier tube; Its solubility is small, not deliquescent, easy to process and use. These advantages make it suitable for on-line measurement of (n, γ) reaction cross sections based on the time-of-flight method.

A closed spherical shell with an inner radius of 10 cm and a thickness of 15 cm can be formed by 42 pentagonal frustums and hexagonal frustums. A BaF$_2$ crystal (end face diameter 14 cm, height 15 cm) is coupled with a 5 in photomultiplier tube and packaged into an independent BaF$_2$ detection unit (Fig. 1(a)). In order to facilitate the entry and exit of the neutron beam, two positions in and out of the GTAF beam line are vacant. The remaining 40 positions are filled with BaF$_2$ detection units, which make up the GTAF (Fig. 1(b)). In the on-line experiment, the neutron beam reacts with the sample in the GTAF center to form a compound nucleus. The compound nucleus in the excited state is deexcited by emitting a cascade of gamma rays. Since the GTAF almost covers the 4π solid angle, the deexcited cascade gamma rays will be detected by the BaF$_2$ detection unit and recorded as (n, γ) reaction events. At the same time, the time of source neutron generation and the time of cascade gamma ray arrival at the detector are recorded to calculate the neutron flight time of (n, γ) reaction and determine the neutron energy[27–30].

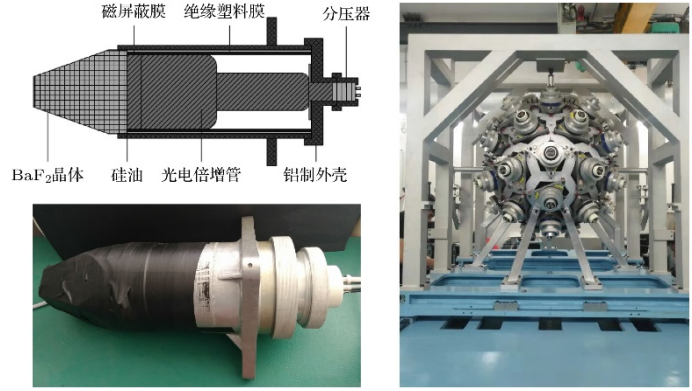

**Figure 1.** (a) BaF$_2$ detector unit; (b) GTAF.

Because Ba and Ra are in the same group, the natural radioactive background of BaF$_2$ crystal is mainly composed of four kinds of α particles in the decay chain of $^{226}$Ra. α particles with energy of 4.7 MeV are emitted from the decay of $^{226}$Ra to $^{222}$Rn, α particles with energy of 5.5 MeV are emitted from the decay of $^{222}$Rn to $^{218}$Po, α particle with energy of 6.0 MeV is produced by the decay from $^{218}$Po to $^{214}$Pb, and α particle with energy of 7.7 MeV is produced by the decay from $^{214}$Po to $^{210}$Pb.

## 3. Selection of different sampling lengths

In this paper, a BaF$_2$ detector unit of GTAF is selected, and a CAEN digitizer DT5751 (10 bit, 1 GS/s) is used to obtain the signal waveform of the detector, and to measure the gamma rays of three radioactive sources ($^{22}$Na, $^{137}$Cs and $^{60}$Co). As shown in the Fig. 2, there are three main types of signals in the BaF$_2$ detection unit: 1) electronic noise, with a fast component and almost no slow component, which comes from the dark current of the photomultiplier tube and other electronic circuits; 2) γ-ray, the effective signal of on-line experiment, has obvious fast and slow components; 3) The α particle of BaF$_2$ crystal itself has almost no fast component and only a slow component.

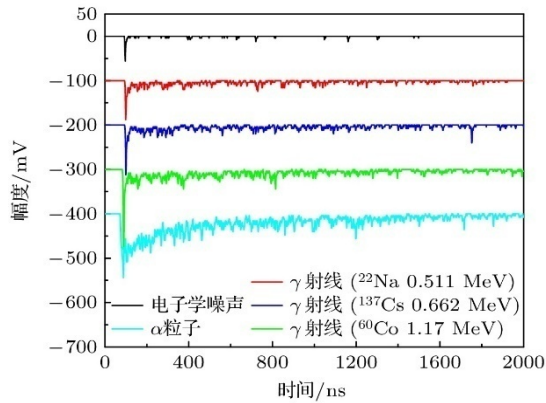

**Figure 2.** Signal waveform of the BaF$_2$ detection unit.

The parameters set by the data acquisition system are: trigger position 100 ns, threshold 20

mV, baseline 0 — 50 ns. The energy spectrum is obtained by integrating the pulse amplitude of the signal waveform. As shown in Fig. 3, the full energy peaks of $^{22}$Na (0.511 MeV), $^{137}$Cs (0.662 MeV), $^{60}$Co (1.17/1.33 MeV) and alpha particles of four energies can be clearly seen. After deducting the background, the energy resolution is obtained by fitting calculation, in which the two full-energy peaks of $^{60}$Co can not be completely distinguished, and is calculated according to the average value of 1.25 MeV.

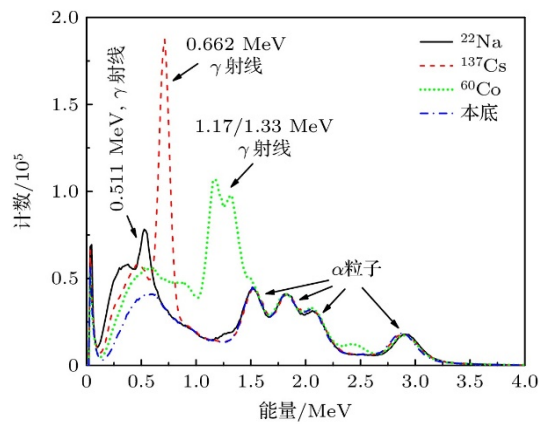

**Figure 3.** Energy spectrum of the BaF$_2$ detection unit.

The Fig. 4 is the comparison of the energy resolution of three kinds of radioactive sources measured by BaF$_2$ detection unit under different integration lengths (the trigger position is the starting point of integration). The energy resolution of the detector unit becomes better with the increase of the integration length from 200 ns, and the energy resolution reaches the best when the integration length is between 1900 ns and 2000 ns, and then the energy resolution becomes worse with the increase of the integration length. Because the light decay time of the slow component of the BaF$_2$ crystal is 620 ns, the waveform pulse basically returns to the baseline at 1900 — 2000 ns, and the complete waveform of the γ-ray signal is just collected at this time, so the energy resolution is the best, so the integration length of the energy spectrum is selected as 2000 ns in the online experiment.

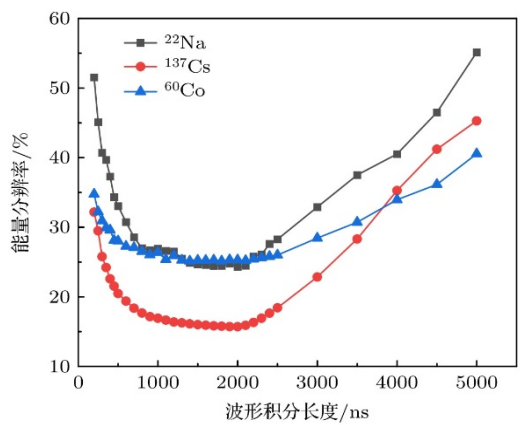

**Figure 4.** Comparison of energy resolution of different integral length.

## 4. α/γ discrimination method

In the current production process, it is difficult to use chemical methods to remove the Ra impurity in the raw material of BaF$_2$ crystal, so it is necessary to use particle identification methods to eliminate the interference of α particles in the BaF$_2$ detection unit on the γ-ray signal as far as possible, so as to effectively reduce the background of on-line experiments.

As shown in Fig. 5, the distribution of waveform eigenvalues of alpha particles and gamma rays can be obtained by using the particle identification method. According to the principle of statistics, the waveform eigenvalues of these two types of particles meet the Gaussian distribution. The discrimination quality factor (FOM) is usually used as the standard to evaluate the particle discrimination capability of the detector in the world. The larger the FOM is, the stronger the discrimination capability is and the better the discrimination effect is.

$$\text{FOM} = \frac{|\mu_\alpha - \mu_\gamma|}{\text{FWHM}_\alpha + \text{FWHM}_\gamma}, \tag{1}$$

Where $\mu_\alpha$ and $\mu_\gamma$ are the mean values of αpeak and γpeak obtained after Gaussian fitting, respectively; FWHM$_\alpha$ and FWHM$_\gamma$ are the full width at half maximum of α peak and γ peak, respectively. The three particle identification methods used in this paper evaluate the identification effect by calculating the FOM value.

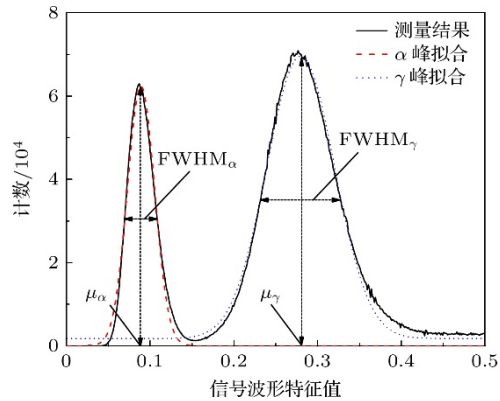

**Figure 5.** Distribution of FOM calculation.

4.1 Fast total component ratio identification method

As shown in the Fig. 6, the γ-ray signal waveform detected by the BaF$_2$ detection unit has an obvious fast component, and then the signal amplitude decreases rapidly and overlaps with the slow component; The signal waveform of alpha particles has no fast component, only slow component, and the signal amplitude decreases slowly. The fast total component ratio discrimination method starts from the signal trigger position, sets different integration lengths to obtain the fast component and the total component of each signal, and calculates the ratio of the fast component and the total component as the waveform characteristic value to obtain the FOM.

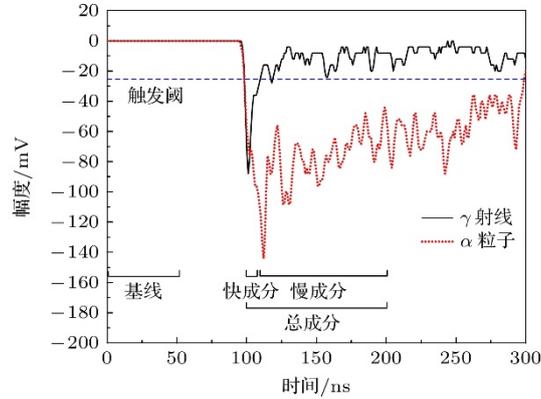

**Figure 6.** Fast and total components of waveform.

In this paper, three kinds of radioactive sources are measured by BaF₂ detection unit, and the FOM of the fast total component ratio of the waveform signal is calculated. The abscissa of the Fig. 7 is the integration length of the total component (50-2000 ns), and F5-F30 represent the integration length of the fast component (5-30 ns), respectively. It can be clearly seen that the longer the integral of the fast component, the smaller the FOM; When the integration length of the total component is 200 ns, the FOM is the largest, so the ratio of the fast component 5 ns to the total component 200 ns is selected for αγ discrimination.

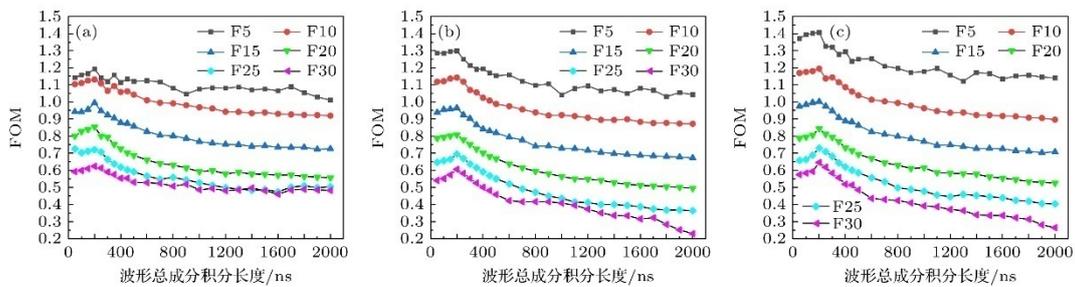

**Figure 7.** FOM with different ratios of fast to total component: (a) $^{22}$Na; (b) $^{137}$Cs; (c) $^{60}$Co.

After the parameters of the fast-to-total component ratio were determined, processing yielded the two-dimensional spectrum shown in Fig. 8. The part circled in the red box is alpha particles (energy 1 — 4 MeV, fast total composition ratio 0.021 — 0.128), which can be clearly distinguished from gamma rays. During the online experimental data processing, the waveform signals in this range will be removed.

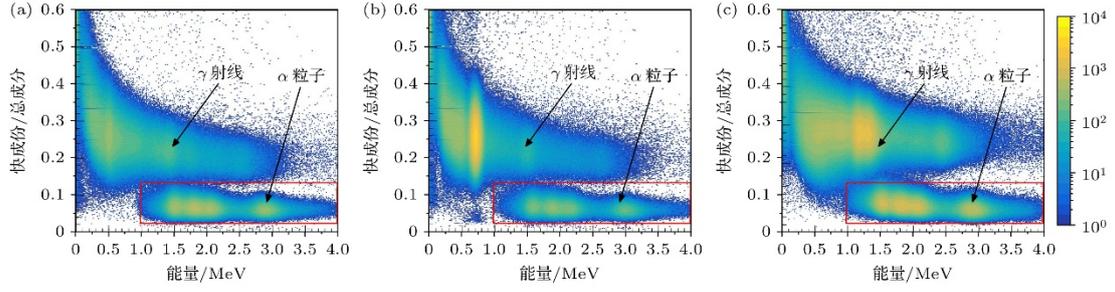

**Figure 8.** Energy versus ratio of fast to total component spectrum: (a) $^{22}$Na; (b) $^{137}$Cs; (c) $^{60}$Co.

## 4.2 Pulse width discrimination

By analyzing the signal waveforms of γ-ray and α particles in BaF$_2$ detection unit, it is found that the pulse widths of the two waveforms are obviously different, which can be used to identify particles. The Fig. 9 is a waveform signal measured by the BaF$_2$ detection unit. First, the peak value of the pulse is obtained through waveform analysis, and then the time to reach 10% of the peak value during the pulse rise is obtained by interpolation calculation$T\_\text{start}_{10\%}$. Similarly, the time to reach 10% of the peak value during the pulse falling edge, denoted as $T\_\text{end}_{10\%}$, was obtained using interpolation.

$$w1 = T\_\text{end}_{10\{\text{\%}\}} - T\_\text{start}_{10\{\text{\%}\}} \, , \tag{2}$$

In the formula, $w1$—$w5$ respectively represent the pulse width of the 10% — 50% peak value of the signal waveform.

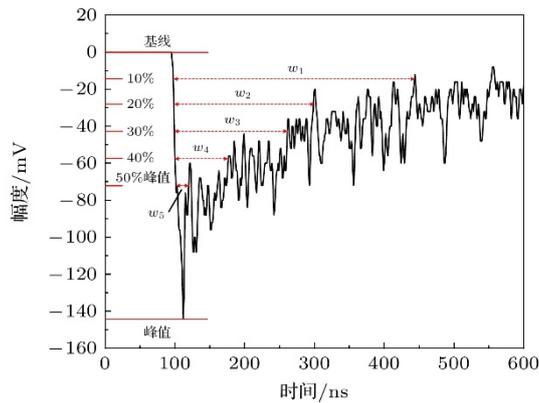

**Figure 9.** Pulse width of waveform.

Using different waveform pulse widths as characteristic values, the FOM values of three radioactive sources are obtained, as shown in Fig. 10, in which the pulse width of 10% peak has the best particle discrimination ability. Because alpha particles have no slow component, compared with gamma rays, the lower the peak ratio is, the longer the pulse width at the corresponding position is, and the easier it is to distinguish them. The Fig. 11 is the two-dimensional spectrum of 10% peak pulse width, and the part circled in the red box is the alpha particle (energy 1-4 MeV, 10% peak pulse width 120-880 ns), which needs to be removed in the data processing.

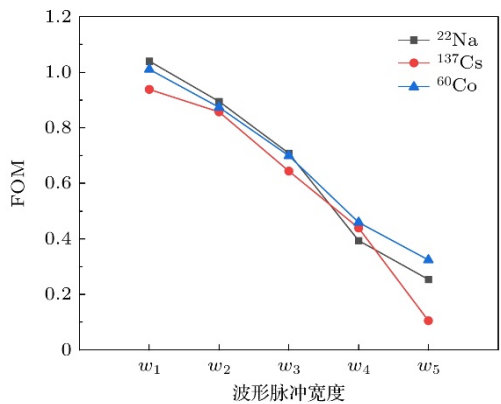

**Figure 10.** FOM with different pulse width of waveforms.

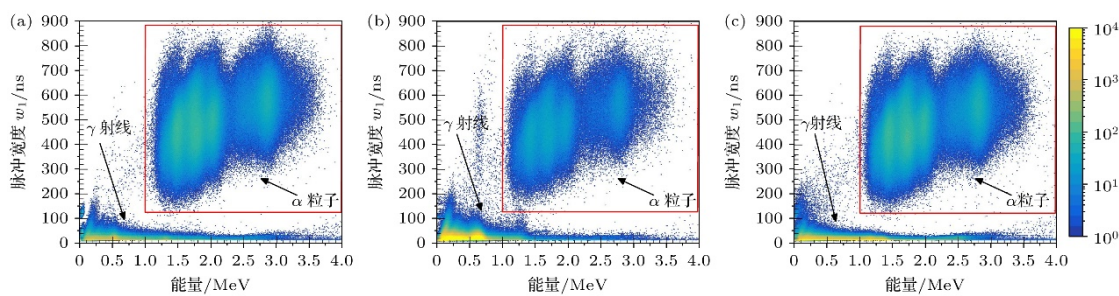

**Figure 11.** Energy versus pulse width spectrum: (a) $^{22}$Na; (b) $^{137}$Cs; (c) $^{60}$Co.

### 4.3 Time decay constant discrimination method

The characteristic of the scintillation light of the BaF$_2$ detection unit is that after the signal waveform reaches the peak position, it decays according to the negative e exponential law, and the waveform amplitude can be fitted by using the formula (3):

$$V_{(t)} = V_0 + A\mathrm{e}^{-(t-t_0)/\tau},\qquad(3)$$

Where $V_{(t)}$ is the amplitude of the waveform at the time of $t$, $V_0$ is the amplitude of the signal baseline, $A$ is the amplitude at the peak of the waveform, $t_0$ is the time at which the peak is reached, and $\tau$ is the time decay constant of the signal waveform.

The waveform of γ-ray signal contains fast component and slow component. After the waveform rises to the peak position rapidly, it will fall to the position of slow component rapidly, and then fall to the baseline position slowly. Therefore, the time attenuation constant $\tau$ of γ-ray signal waveform is mainly determined by the fast component. The signal waveform of α particle has almost no fast component but only slow component, and its time decay constant $\tau$ is mainly determined by the slow component. The decay time of the fast component of the BaF$_2$ detection unit is 0. 6 ns, and the decay time of the slow component is 620 ns. The difference between the two is obvious, which can be used to identify alpha particles and gamma rays. As shown in the Fig. 12, the peak value of the pulse was obtained by waveform analysis, and the waveform 2000 ns after the peak position was selected, and the time decay constant was obtained by fitting with the formula (3).

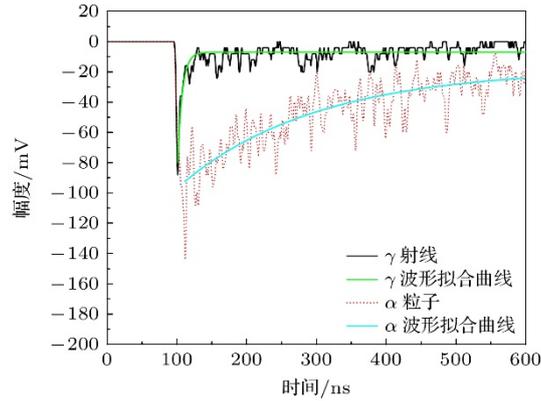

**Figure 12.** Fitting of the time decay constant of waveform.

The Fig. 13 shows the two-dimensional spectrum of the time decay constant, and the alpha particle is the part circled in the red box (the energy is 1-4 MeV, and the time decay constant is 55-45 s$^{-1}$). It can be seen that in the range of 1-1.3 MeV, there is a partial overlap between the gamma ray and the alpha particle, and it cannot be clearly distinguished.

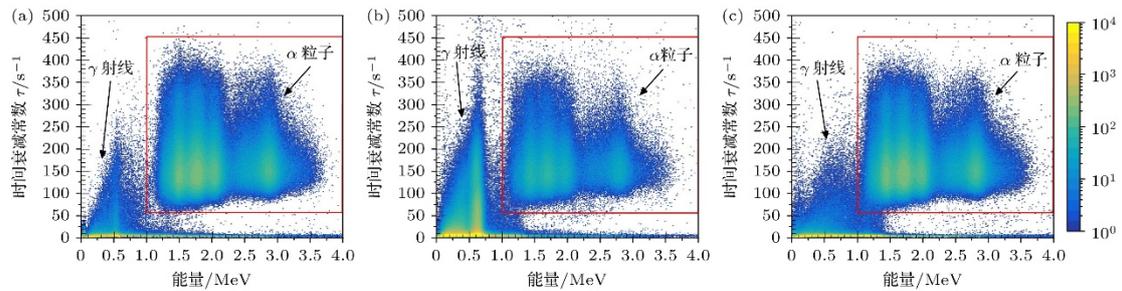

**Figure 13.** Energy versus time decay constant spectrum: (a) $^{22}$Na; (b) $^{137}$Cs; (c) $^{60}$Co.

## 5. Analysis and discussion of experimental result

The optimal setting of FOM among the three particle identification methods is selected, and the comparison results are shown in Tab. 1. The same method is applied to different radioactive sources, and the FOM values obtained are not very different, which proves that each method is suitable for on-line experimental measurement to distinguish gamma rays with different energies from alpha particles. The FOM values of the pulse width discrimination method and the time decay constant discrimination method are lower than those of the fast composition ratio discrimination method. The main reason is that the signal waveform of the large volume BaF$_2$ detection unit has obvious fluctuations (many burrs appear as shown by Fig. 9 and Fig. 12), which will seriously affect the statistical distribution of the characteristic value calculation results when calculating the pulse width and fitting the decay constant, resulting in larger FWHM values of α peak and γ peak, thus reducing the FOM value.

**Table 1.** Comparison of FOM of different particle identification methods.

| Radioactive source | Fast total component ratio (fast component UNK05 ns/total component 200 ns) | Pulse width (10% peak)/ns | Time decay constant/s$^{-1}$ |
|---|---|---|---|
| $^{22}$Na | 1.19 | 1.04 | 1.07 |
| $^{137}$Cs | 1.30 | 0.94 | 0.93 |
| $^{60}$Co | 1.41 | 1.01 | 0.96 |

After using the three particle identification methods, it can be seen from the energy spectrum of the radioactive source that the α particles are significantly reduced, and the full energy peak (Fig. 14(a)) of $^{22}$Na (1.27 MeV), which is submerged by the α particle count, and the 2.5 MeV γ peak (Fig. 14(c)), which is produced by the two cascade γ rays of $^{60}$Co detected by the same BaF$_2$ detection unit, are also exhibited. From the results, the discrimination effect of alpha particles and gamma rays with fast total component ratio (fast component 5 ns, total component 200 ns) is the best, and the discrimination effect of pulse width (10% peak value) and time decay constant is not very different, which is also consistent with the calculation results of FOM value.

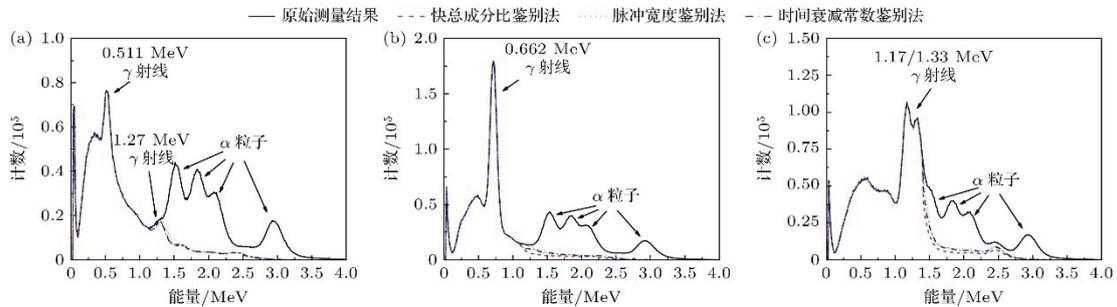

**Figure 14.** Comparison of energy spectra obtained by different particle identification methods: (a) $^{22}$Na; (b) $^{137}$Cs; (c) $^{60}$Co.

## 6. Conclusion

Based on the γ-ray measurement results of the large volume $BaF_2$ detector unit, the influence of the sampling length of the signal waveform on the energy resolution is studied, and the best energy resolution is obtained when the sampling length is determined to be 2000 ns. According to the signal waveform characteristics of $BaF_2$ detection unit, the discrimination ability of α particles and γ rays was studied by using three methods of fast composition ratio, pulse width and time decay constant, and the quality factor FOM was used as the evaluation value. By comparing the energy spectrum of radioactive source and FOM value, the α/γ discrimination ability of fast total component ratio is better than the other two methods.

At present, the GTAF, which is composed of 40 $BaF_2$ detection units, carries out the (n, γ) reaction cross section on-line measurement experiment. The data acquisition system adopts the signal full waveform acquisition mode. The amount of data recorded, transmitted and stored in each experiment is very large, which is not conducive to data storage and processing. At the same time, it also prolongs the dead time of the data acquisition system and affects the uncertainty of the cross section data. In this paper, the advantages of the 2000 ns sampling length and the fast total component ratio identification method are verified. The next step is to upgrade the data acquisition system of the online experiment, which only records a series of parameters such as the threshold crossing time of each signal waveform (for the time-of-flight method), the 5 ns amplitude integral value after threshold crossing (fast component), the 200 ns amplitude integral value (total component), the 2000 ns amplitude integrated value (energy), and the detection unit number. Without recording the full waveform, the use of these information enables online data processing for both alpha particle background rejection and (n, γ) reaction identification. It is estimated that the amount of data stored in the upgraded acquisition system is reduced from 118 MB/s to 24 MB/s, which can significantly reduce the dead time of the data acquisition system and improve the accuracy of the cross section data.